\documentclass{emulateapj}

\slugcomment{}

\shorttitle{Spatial Periodicities in a Coronal Loop}
\shortauthors{D.B. Jess et al.}

\begin{document}

\title{Discovery of Spatial Periodicities in a Coronal Loop using Automated 
Edge-Tracking Algorithms}

\author{D. B. Jess}
\affil{Astrophysics Research Centre, School of Mathematics and Physics, Queen's University, Belfast, BT7~1NN,
Northern Ireland, U.K.}
\affil{}
\affil{NASA Goddard Space Flight Center, Solar Physics Laboratory, Code 671, Greenbelt, MD 20771, USA}
\email{djess01@qub.ac.uk}

\author{M. Mathioudakis}
\affil{Astrophysics Research Centre, School of Mathematics and Physics, Queen's University, Belfast, BT7~1NN,
Northern Ireland, U.K.}

\author{R. Erd\'{e}lyi and G. Verth}
\affil{SP$^{2}$RC, Department of Applied Mathematics, The University of Sheffield, Sheffield, S3 7RH,
England, U.K.}

\author{R. T. J. McAteer}
\affil{NASA Goddard Space Flight Center, Solar Physics Laboratory, Code 671, Greenbelt, MD 20771, USA}
\affil{}
\affil{Catholic University of America, 620 Michigan Ave., N.E. Washington, DC 20064, USA}

\and

\author{F. P. Keenan}
\affil{Astrophysics Research Centre, School of Mathematics and Physics, Queen's University, Belfast, BT7~1NN,
Northern Ireland, U.K.}

\author{~}
\affil{~}

\begin{abstract}
A new method for automated coronal loop tracking, in both spatial 
and temporal domains, is presented. Applying this technique to 
{\sc{trace}} data, obtained using the 171\AA~filter on 1998 July 14, 
we detect a coronal loop undergoing a 270~s kink-mode oscillation, 
as previously found by 
Aschwanden~et~al.~\cite{Asc99}. However, we also detect 
flare-induced, and previously unnoticed, {\it spatial 
periodicities} on a scale of 3500~km, which occur along the 
coronal-loop edge. Furthermore, we 
establish a reduction in oscillatory power for these spatial 
periodicities of 45\% over a 222~s interval. We relate the 
reduction in detected oscillatory power to the physical damping of 
these loop-top oscillations.
\end{abstract}

\keywords{Methods: data analysis --- Techniques: image processing
--- Sun: corona --- Sun: evolution --- Sun: oscillations}

\section{Introduction}
\label{intro}

Automated feature recognition and tracking has long been a goal for 
scientists. With the advent of higher sensitivity satellite-based 
telescopes and the resulting increase in readout rates, it is 
imperative to be able to do some preliminary data analysis 
in real time. Without this ability, the large data rates 
achievable make it extremely difficult to implement onboard trigger 
programmes (e.g. a flare trigger) which allow the telescope 
pointing to be redirected to an area of interest almost immediately. 
Furthermore, having a means of establishing solar phenomena 
in real time allows instrument users to select 
datasets of particular interest much more readily than in 
the past. Thus, many schemes have been set up to push this form of 
real-time data analysis forward, including the Heliophysics 
Knowledge Base programme, which will enable real-time 
detections of oscillatory phenomena occurring on the Sun.

\begin{figure*}
\epsscale{1.0}
\plotone{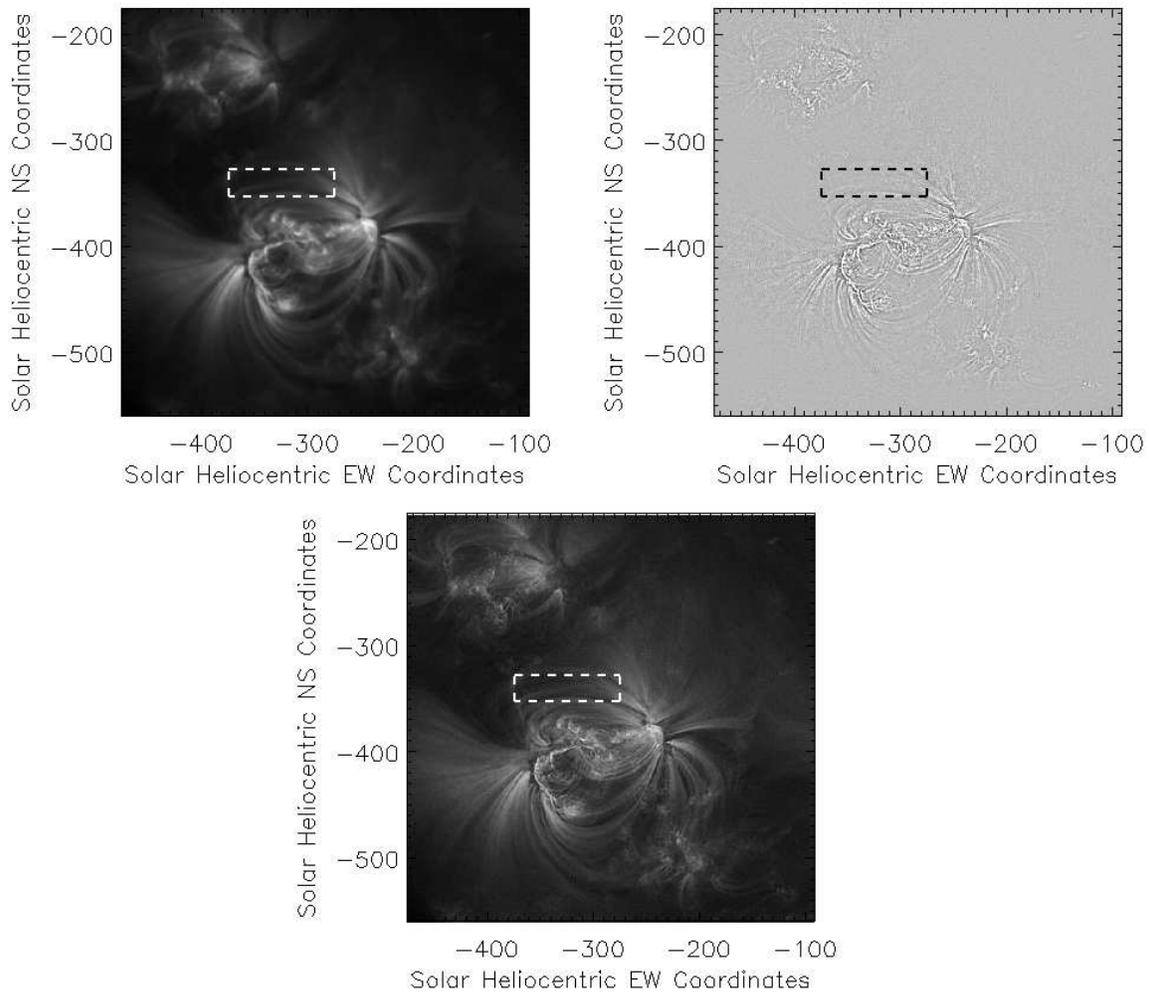}
\caption{Complete {\sc{trace}} field of view (top left) after undergoing initial processing using the {\sc{trace\_prep.pro idl}}
routine. The top right image shows the fine-scale structures detected using the Laplacian filter outlined in \S~\ref{Laplace} and
the bottom image reveals the sharpened {\sc{trace}} field of view after the addition of fine-scale structures. Note how the clarity
of loop structures is drastically increased. Overplotted dashed 
lines in each image outlines the region of interest selected for this paper. \label{TRACEfull}}
\end{figure*}

Current satellites, in particular the {\it{Transition 
Region And Coronal Explorer}}, {\sc{trace}}, have enabled the 
analysis of many oscillatory signatures originating within the outer 
solar atmosphere. Aschwanden~et~al.~\cite{Asc99}, 
Schrijver~et~al.~\cite{Sch02}, Wang \&~Solanki~\cite{Wan04}, 
De~Pontieu~et~al.~\cite{Dep05} and Marsh~\&~Walsh~\cite{Mar06}, just 
to name a few typical examples (a recent review is in e.g. 
Banerjee~et~al.~2007), all report oscillatory phenomena 
occurring within coronal loop structures. Previously, 
however, the majority of work on oscillating coronal loops has been 
focused on the temporal domain (e.g., Nakariakov~et~al.~1999). 
However, it was suggested by Erd\'elyi~\&~Verth~\cite{erd07} and 
Verth~et~al.~\cite{ver07} that the imaging capacities of current 
and future high-resolution instruments should allow us to investigate 
oscillatory phenomena in the spatial domain (e.g. along the observed 
waveguide). Inspired by these suggestions, in this work we 
investigate the spatial evolution of a coronal loop structure, 
independent of its position during a flare-induced kink oscillation. 
From this we can understand the behaviour of the loop, in the 
spatial domain, and analyse any phenomena which are flare induced, 
yet confined to the volume occupied by the loop. In 
\S~\ref{observations} we describe the observations, while in 
\S~\ref{analy} we discuss the methodologies 
used during the analysis of the data which allow us to accurately 
track the spatial evolution of coronal loops. A discussion of our 
results in the context of confined loop oscillations is given in 
\S~\ref{results}, and our concluding remarks are in 
\S~\ref{conc}.

\section{Observations}
\label{observations}

The data presented here are part of an observing sequence obtained 
on 1998 July 14, using the {\sc{trace}} imaging satellite. The 
optical setup of {\sc{trace}} allowed a $384\arcsec~\times~384\arcsec$~area 
surrounding active region NOAA~8270 to be 
investigated with a spatial sampling of $0.5\arcsec$~per pixel. 
The 171\AA~filter was selected for these observations, 
which has an inherent passband width of 6.4\AA, 
allowing plasma in the temperature range 0.2--2.0~MK to be studied. 
The cadence of the {\sc{trace}} instrument was not constant during 
the observing sequence, with the time between successive exposures 
varying between 66 and 82~s.

Our dataset consists of 88 
successive {\sc{trace}} images providing nearly two hours of 
continuous, uninterrupted data. During the observing sequence, a 
large M4.6 flare occurred in the immediate vicinity of the active 
region under investigation.

\section{Data Analysis}
\label{analy}

The {\sc{trace}} data was retrieved directly from the Lockheed 
Martin Solar and Astrophysics Labs database and was subjected to 
standard processing algorithms: The {\sc{idl}} 
routine {\sc{trace\_prep.pro}} was implemented to remove 
cosmic-ray streaks, reduce readout noise and 
prepare the data in a user-friendly format. De-rotation of the 
{\sc{trace}} images was also deemed necessary, since the observing 
sequence lasts approximately two hours, equating to a relatively 
large solar rotation ($20\arcsec$ at disk centre). From this point,
it was possible to use additional programmes to aid in the analysis
of the data as described in detail below.

\begin{figure}
\epsscale{1.0}
\plotone{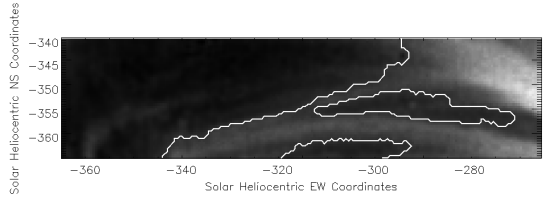}
\caption{A contoured (bold white line) {\sc{trace}} coronal loop using a minimum intensity threshold of the 
background median value plus 5~sigma. From this threshold value we can see that the loop is contoured accurately 
with minimal solar background included. \label{contoured}}
\end{figure}

\begin{figure}
\epsscale{1.0}
\plotone{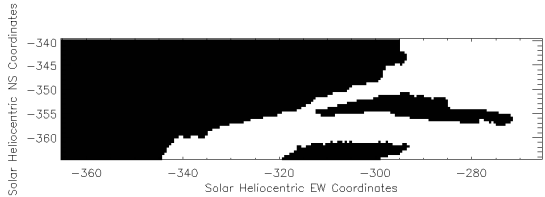}
\caption{The contours obtained in Figure~\ref{contoured} are converted to a binary scale as defined in \S~\ref{Fractal}. This provides
the sharp intensity gradient required for accurate feature edge detection and tracking. \label{binary}}
\end{figure}

\subsection{Laplacian Image Sharpening}
\label{Laplace}

We implement a form of Laplacian 
image sharpening. Using the same 3-by-3 convolution kernel implemented 
by Gonzalez \& Woods~\cite{Gon92}, we are able to enhance the fine-scale 
structure found in the {\sc{trace}} images. Figure~\ref{TRACEfull} shows the 
result of applying the convolution kernel to a full {\sc{trace}} image. 
Note the amount of fine detail found in this filtered image (top-right 
panel of Fig~\ref{TRACEfull}). The 
next step is to add this fine-scale information back into the 
original image in order to sharpen the image. Note, in the 
bottom panel of Figure~\ref{TRACEfull}, the emphasis of 
fine structures when compared to the original, unfiltered image. 
Figure~\ref{TRACEfull} demonstrates that 
these processes are readily applicable to full-size 
{\sc{trace}} fields of view due to the fast processing ability of 
modern day computers. For the purposes of the data under 
investigation here, all {\sc{trace}} images were subjected to 
Laplacian image sharpening routines.

\subsection{Wavelet Modulous Maxima Edge Detection}
\label{Fractal}

After successful completion of image sharpening, edge detection 
algorithms can be implemented. The sharpened images have increased 
fine-scale information and as such provide the necessary platform 
for establishing feature edges which may have been previously 
unresolvable. Immediately after the flare event, one particular 
coronal loop is seen to oscillate. This kink-mode oscillation 
has been investigated many times (see Aschwanden~et~al.~1999 for the 
initial investigation) in the temporal domain. However, following 
Erd\'elyi~\&~Verth~\cite{erd07}, we propose here to investigate the 
behaviour of the loop in the spatial domain. In order to track the 
spatial behaviour of the loop, an automated routine was devised 
to minimise errors introduced through human interaction with the 
data.

The first step involves intensity thresholding, whereby features in 
the {\sc{trace}} field-of-view are contoured depending on their 
emissive flux. Threshold limits for contouring are entirely 
arbitrary and as such may vary from dataset to dataset as well 
as for which features are desirable to contour. However, for the data 
presented here, a lower intensity threshold of the background median 
value plus 5~sigma is used. This value allowed us to contour 
{\sc{trace}} coronal loop structures, yet leave out background quiet 
Sun as shown in Figure~\ref{contoured}. To emphasize feature edges 
and remove shallow intensity gradients, a binary format for feature 
mapping is used. All pixels which lie above the lower intensity 
threshold defined above are assigned a value of `1'. Those pixels 
which lie below the threshold were assigned a value of `0', as seen 
in Figure~\ref{binary}. This binary format provides an absolute 
intensity cut-off thus providing a definite feature edge which can 
be tracked, both spatially and temporally. Placing the images into 
binary format was deemed necessary due to the poor spatial resolution of the 
{\sc{trace}} spacecraft, whereby neighbouring bright features are hard to 
distinguish unless a form of thresholding is used. 
Future instruments, which provide higher spatial resolution, should eradicate 
the need for binary formatting. The binary image is 
padded using a zeroed array from which feature edge tracking can 
commence.

\begin{figure}
\epsscale{1.0}
\plotone{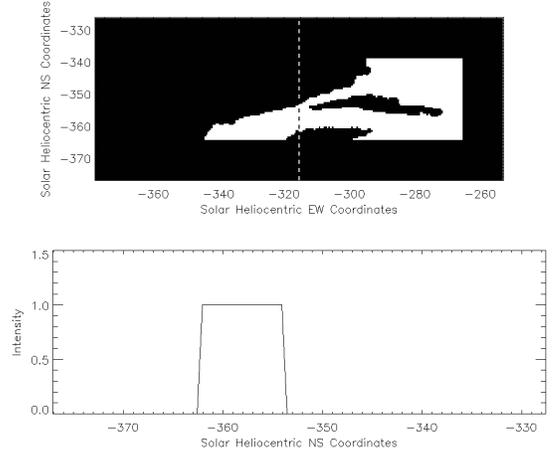}
\caption{Upper panel shows the binary-formatted image obtained in Figure~\ref{binary} inserted into a padded, zeroed array 
which provides the best platform for WMM edge detection. A vertical dashed line at x pixel 125, corresponding to a 
Heliocentric E/W coordinate of -315$''$, is drawn, with the resulting intensity plot, along the N/S direction for this x pixel, 
shown in the bottom panel. \label{puffed_yline}}
\end{figure}

\begin{figure}
\epsscale{1.0}
\plotone{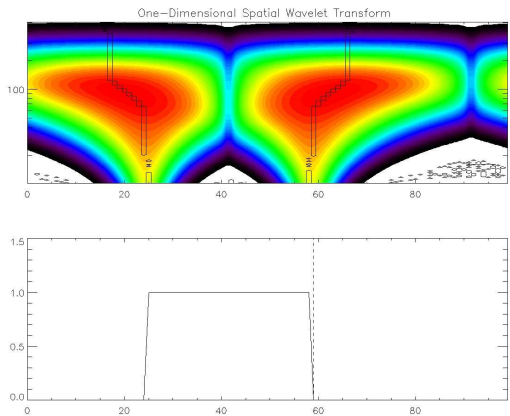}
\caption{The wavelet power spectrum (top) generated along the vertical spatial direction of the {\sc{trace}} field of view
for horizontal pixel number 125. This panel also shows regions of maximum power being traced down to the intersect point on the x-axis.
The intensity plot from Figure~\ref{puffed_yline} is again plotted (bottom) with a dashed line indicating the position of intersection
of the maximum wavelet power traces with the x-axis. Notice how the x-axis intersect position of maximum wavelet power corresponds
exactly to the reduction of intensity in the bottom panel. This reduction in intensity is analagous with the upper edge of the
{\sc{trace}} coronal loop showing the accuracy of this method when detecting feature edges. \label{fractal_x125}}
\end{figure}

Tracking of this feature edge is performed using a Wavelet Modulous 
Maxima (WMM) technique (Muzy~et~al.~1993, McAteer~et~al.~2007). From 
the padded binary array created above, a horizontal pixel is chosen. 
Figure~\ref{puffed_yline} shows the selection of horizontal pixel 
125, as indicated by the vertical dashed line, as well as the 
corresponding intensity plot along this vertical line. A wavelet 
transform is produced for this intensity plot, demonstrated in 
Figure~\ref{puffed_yline}, utilizing a Mexican Hat wavelet. A 
Mexican Hat wavelet, which is a double derivative of the traditional 
Morlet wavelet, is very useful for detecting sharp intensity 
gradients which are present due to the binary format implemented 
above. An absolute wavelet power spectrum is plotted as a function 
of vertical pixel element in Figure~\ref{fractal_x125}. 
Additionally, maximum wavelet power features are traced down to the 
x-axis of the power spectrum and reveal the locations of maximum 
intensity gradients. Figure~\ref{fractal_x125} re-plots the 
intensity in the vertical direction at horizontal pixel 125 and it 
can be seen that where the maximum wavelet power meets the x-axis 
also corresponds to the edges of the {\sc{trace}} coronal loop under 
investigation.

From Figure~\ref{contoured}, it is clear that better 
contrast is achieved at the Northern edge of the loop where its 
brightness overlies fainter, less intense, quiet Sun. 
By contrast, the Southern edge of the loop in Figure~\ref{contoured} 
is difficult to separate from other loop structures, due to the 
densely packed nature of such features closer to active regions. 
Since in this instance it is desirable to track the upper edge of 
the coronal loop, we disregard the first maximum wavelet power 
location and record the position where the second maximum wavelet 
power location intersects with the x-axis. This pin points the exact 
location of the Northerly edge of the coronal loop in question. This 
process is repeated over the entire E/W direction, and for all 
frames in the observing sequence, with the subsequent position of 
the Northern edge of the {\sc{trace}} coronal loop recorded. From 
the resulting loop edge positions, it is possible to analyse both 
temporal, and spatial, variations.

\section{Results and Discussion}
\label{results}

\begin{figure}
\epsscale{1.0}
\plotone{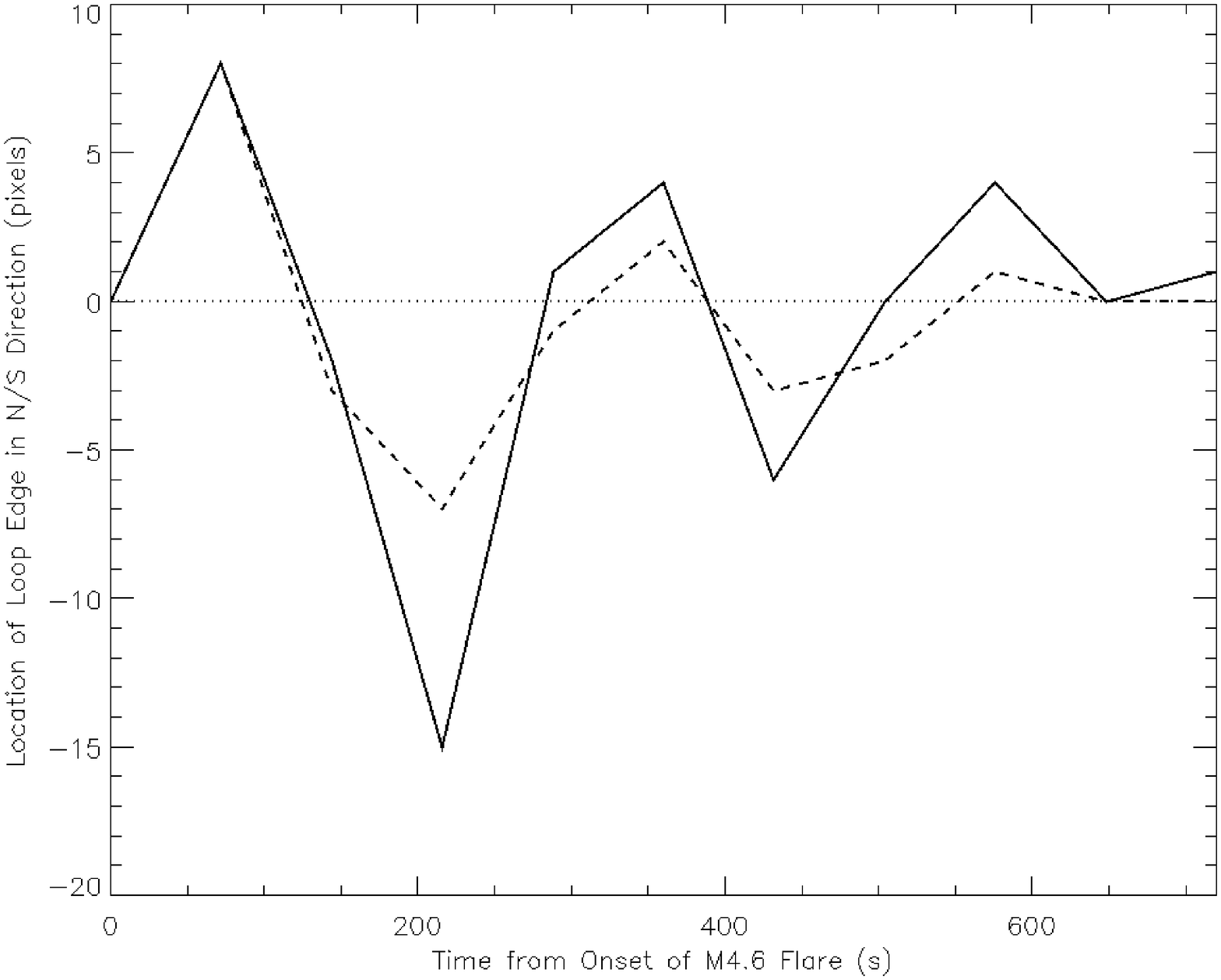}
\caption{The temporal evolution of a particular portion of the oscillating coronal loop. Here, shown as a bold line, we see a 
periodicity of approximately 270~s related to the kink-mode oscillation of the loop and the subsequent damping of the 
amplitude with time. The dashed line indicates the temporal periodicity detected when a lower-intensity threshold is used to 
contour the loop edge, while the dotted line indicates the equilibrium position of the loop. A lower intensity threshold 
(see \S~\ref{results}) reveals the same periodicity, indicating that the loop-edge 
tracking, due to high contrast between the loop and surrounding quiet sun, is not overly sensitive to threshold values.
\label{TRACEtime}}
\end{figure}

Our main goal is to investigate the spatial behaviour of the 
oscillating coronal loop. However, 
to verify that the presented loop-tracking routine is functioning 
properly, derived results in the temporal domain are compared to 
previous findings. To compensate for the timing irregularity 
outlined in \S~\ref{observations}, all data were interpolated onto 
a constant-cadence time series with linear interpolation performed 
between data points. This provides the necessary platform for 
temporal studies to commence. Maintaining consistency with the previous 
example, a plot indicating the variation with time of the location 
of the Northern edge of the coronal loop, for horizontal pixel~125, 
is displayed in Figure~\ref{TRACEtime}. It is clear that a periodicity of 
approximately 270~s exists, consistent with the findings of 
Aschwanden~et~al.~\cite{Asc99}. This result implies that the loop 
detection and tracking algorithms, developed here, are functioning 
accurately.

\begin{figure}
\epsscale{1.0}
\plotone{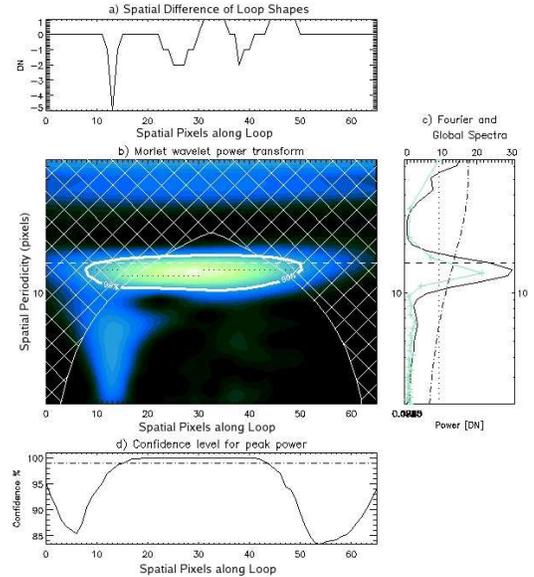}
\caption{The detection of a 10~pixel spatial oscillation. The resulting Northern loop-edge outline created by the subtraction
of the loop edge at 0~s from the loop edge at 280~s after the flare event is shown in a). The one-dimensional spatial wavelet
power transform along with locations where detected power is at, or above, the 99\% confidence level are contained within the contours
in b). Plot c) shows the summation of the wavelet power transform over time (full line) and the Fast Fourier power spectrum (crosses)
over time, plotted as a function of period. Both methods have detected a well pronounced 10~pixel spatial oscillation. The global wavelet
(dashed dotted line) and Fourier (dotted line) 95\% significance levels are also plotted. The cone of influence (COI), cross-hatched
area in the plot, defines an area in the wavelet diagram where edge effects become important and as such any frequencies outside the
COI are disregarded. Periods above the horizontal line (dotted) fall within the COI. The probability levels, which
are related to the percentage confidence attributed to the peak power at every time step in the wavelet transform, are plotted in d).
\label{subtracted_flare2}}
\end{figure}

From Figure~\ref{TRACEtime}, it appears that the oscillating coronal 
loop passes through an equilibrium position at times of 0, 280 and 
502~s, with the overall shape of the loop similar at the 
corresponding {\sc{trace}} frames. If indeed the shape of the loop 
is identical at times of 0, 280 and 502~s, then a subtraction of any 
two of these loop shapes should provide a resulting image equal to 
zero. However, upon subtraction of the loop shape at 0~s from that 
at 280~s, a spatial periodicity is revealed. Running this resulting 
loop shape through a one-dimensional spatial wavelet transform 
establishes an oscillatory period along the Northern loop edge of 
10~pixels (Fig~\ref{subtracted_flare2}). A number of strict 
criteria implemented on the data, including the test against 
spurious detections of power that may be due to Poisson noise, the 
comparison of the input lightcurve with a large number (1500) of 
randomized time-series with an identical distribution of counts 
(see O'Shea~et~al.~2001 for a detailed explanation) and 
the exclusion of oscillations which last, in duration, less than 
$1.41$ cycles, allowed us to insure that 
oscillatory signatures correspond to real periodicities. These 
criteria have been described in detail in previous papers (see 
Jess~et~al.~2007, Banerjee~et~al.~2001, McAteer~et~al.~2004, 
Ireland~et~al.~1999, Mathioudakis~et~al.~2003). Furthermore, 
according to the Ritz theorem of variational principles (Ritz~1908), spatial 
amplitude dependence is much more sensitive to structural changes of 
the magnetic loop ({\it{e.g.}} stratification, magnetic field 
divergence, twist etc.) than the traditionally studied period ratios 
of FFT-identified modes of oscillations. Thus, analysing spatial 
variations, rather than the more traditional temporal variations, 
will provide a useful platform in which to maximize the limited 
resolution of current satellite imagers. For the {\sc{trace}} plate 
scale equal to $0.5\arcsec$ per pixel, the detected 10~pixel 
periodicity corresponds to an absolute spatial distance, or wavelength, of 
approximately 3500~km per complete oscillation (more than three 
complete cycles detected).

\begin{figure}
\epsscale{1.0}
\plotone{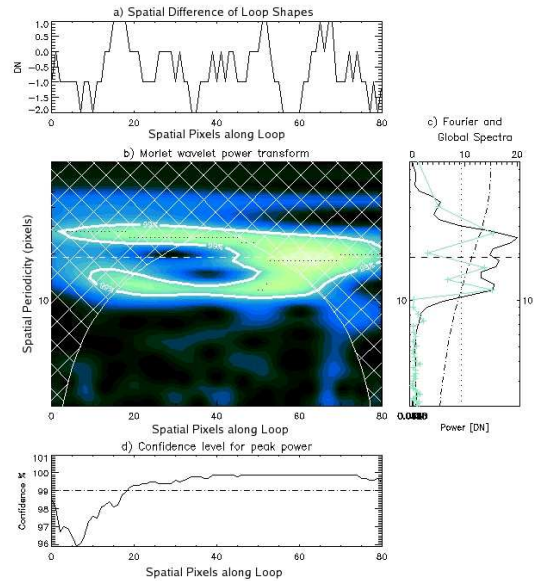}
\caption{Identical plot to Figure~\ref{subtracted_flare2}, except the loop-edge outline shown in a) results from the subtraction of the
loop edge at 0~s from the loop edge at 502~s after the flare event. We can see that a 10~pixel spatial periodicity is still evident,
albeit with a reduction in normalised oscillatory power.
\label{subtracted_flare}}
\end{figure}

To investigate if this periodicity is visible at the following equilibrium 
position (502~s), we subtract the loop shape at 0~s from that at 
502~s, and pass through our one-dimensional spatial wavelet 
transform. The resulting evidence, shown in 
Figure~\ref{subtracted_flare}, indicates that the 10~pixel spatial 
periodicity is indeed still present, albeit with a reduction in 
oscillatory power. We interpret the reduction in oscillatory power 
as the physical signature of damping of the loop-top oscillations. 
On further inspection of 
Figures~\ref{subtracted_flare2}~and~\ref{subtracted_flare}, we 
see that the oscillatory power, in normalised digital number (DN), has dropped 
from 29 to 16 over the 222~s interval. This equates to a 45\% 
decrease in oscillatory power and is consistent with other aspects 
of coronal-loop oscillations, which are seen to be heavily damped 
(Aschwanden~et~al.~2003, Ruderman~2005, Terradas~et~al.~2006, 
Dymova~\&~Ruderman~2007).

\begin{figure}
\epsscale{0.85}
\plotone{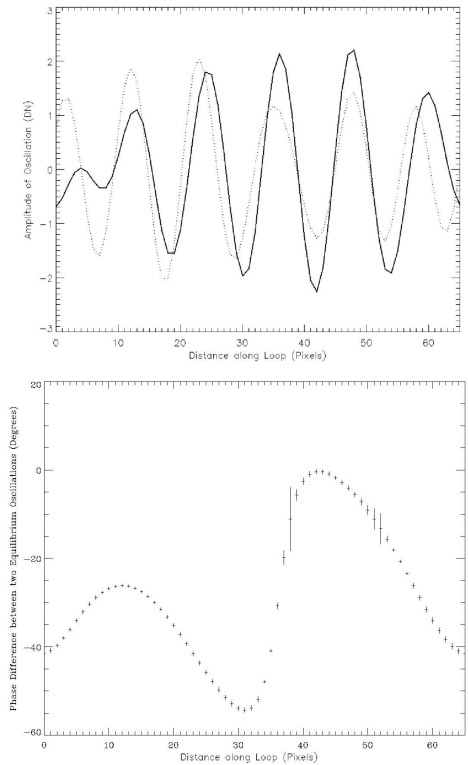}
\caption{The top panel shows the oscillation amplitude as a function of position along the loop for both loop-top oscillations. The solid
line relates to the spatial periodicity detected between times of 0~and~280~s, while the dotted line relates to the spatial periodicity
detected between times of 0~and~502~s. From this plot, it is clear to see a phase shift between the two loop-top oscillations. The bottom
panel shows the phase-difference results of a cross-wavelet spectrum of the two detected spatial oscillations. Maximum phase 
deviations from the average spatial-phase curve, as detailed in \S~\ref{results}, are plotted as vertical error bars. We can clearly 
see a phase shift, on the order of $0^{\circ}$ to $-50^{\circ}$, over the duration of 6~complete cycles.
\label{phase}}
\end{figure}

To investigate a possible phase relationship (Athay~\&~White~1978, 
O'Shea~et~al.~2007), we create a cross-wavelet power spectrum 
between the two detected loop-top oscillations, and inspect the 
resulting spatial-phase diagram (Fig~\ref{phase}). This figure 
reveals a clear phase shift, over many complete wavelengths,  
of $0^{\circ}$ to $-50^{\circ}$. Due  
to the long kink-oscillation period ($\approx$~270~s), and the 
relatively poor, and irregular, cadence of the {\sc{trace}} 
instrument, we are only able to evaluate kink-equilibrium spatial 
periodicities at two times during the presented dataset. At these 
two times, the coronal loop undergoing the kink oscillation has 
returned to it's equilibrium state - i.e. matching the shape and 
position of the loop prior to the flare-induced kink oscillation. 
Between these two equilibrium states, the loop in question has 
undergone one complete kink-oscillation cycle.

However, since we cannot 
evaluate more than two kink-equilibrium spatial periodicities, 
we are unable to determine the direction of wave propagation. Three or more 
kink-equilibrium positions would allow for a comparison of phase differences, 
thus providing an indication to the direction of wave propagation. Thus, a 
phase shift of $0^{\circ}$ to $-50^{\circ}$ may equally be represented as 
$(n360)^{\circ}$ to $(n360 - 50)^{\circ}$, where $n$ is an integer number arising from 
the low sampling rate of the {\sc{trace}} instrument. As described 
by Athay~\&~White~\cite{Ath79} and O'Shea~et~al.~\cite{Osh06}, the errors 
associated with the phase shift can only be constrained and determined 
unambiguously by careful measurements over the full range of observed 
wavelengths. To reduce the errors involved with phase difference analysis we 
evaluate the phase shift over a number of different wavelengths. Through 
examination of the Fourier and global spectra associated with the detected 
3500~km spatial oscillation, in addition to the boundaries of the 99\% confidence-level 
contours, we find that the FWHM of the presented oscillation is 5~pixels, or 
$\approx$~1800~km. This corresponds to 42~wavelength resolution elements, when using 
a 0.016 spacing between discrete scales, in the evaluation of the cross-wavelet 
power spectrum (Torrence~\&~Compo~1998). Performing phase difference analysis on each 
of the 42~wavelength elements, spanning a wavelength coverage of 2600--4400~km, we are 
able to verify the existence of a $0^{\circ}$ 
to $-50^{\circ}$ phase shift. More importantly, we can reduce the error associated with 
the calculated phase differences through examination of the multiple wavelength elements. 
The bottom panel of Figure~\ref{phase} includes the maximum deviation, from the average 
spatial-phase diagram, induced by the analyses 
of 42~wavelength resolution elements. It can be seen from Figure~\ref{phase} that the phase 
error bars do not detract from the generalized shape of the spatial-phase diagram, thus 
validating the existence of a $0^{\circ}$ to $-50^{\circ}$ phase shift.

%If this loop-top oscillation is a propagating wave, using the confines 
%of the coronal loop as a waveguide, then it must exhibit a velocity matching 
%the Alfv\'{e}n speed. Assuming an Alfv\'{e}n speed of $\approx$~1000~km/s 
%(Nakariakov~\& Verwichte~2005), and using the time interval (322~s) between 
%successive equilibrium positions, we can derive a traversed distance for the 
%loop-top wave of $\approx$~0.32~Mm (equivalent to 88.9 periods). 
%Subtracting complete periods from this value (88.9~-~88), the remainder is 
%0.9, which corresponds to a phase of 324$^{\circ}$. This is comparable to the 
%$0^{\circ}$ to $-50^{\circ}$ phase relation shown in Figure~\ref{phase}.

In order to validate our results, we have undertaken a number of further 
rigorous tests. We have implemented the same loop-edge detection and 
tracking processes on additional coronal loops away from the region 
under investigation, yet within the same {\sc{trace}} field of view. 
Since these loops are not connected with the flare event, and 
therefore not seen to oscillate, no spatial periodicities along the 
loop would be expected when two loop outlines are subtracted from 
one and other. Subtracting the control loop outline at 0~s from 
that at 280~s after the flare, to remain consistent with the work 
above, and running through a one-dimensional spatial wavelet 
transform reveals no periodicities. Furthermore, we tested for 
spatial periodicities, on the loop seen to oscillate, both prior to, 
and long after, the flare event. These tests found no oscillatory 
phenomena in either instance, indicating that the spatial 
periodicities detected are flare induced. To further test the 
abilities of the presented algorithm, we carry out an identical form 
of analysis on an additional case of an oscillating coronal loop observed by 
{\sc{trace}} on 2001 April 15. We chose a loop arcade which is seen to 
oscillate after being buffeted by a neighbouring X14.4 flare originating 
from NOAA~9415. Implementing our algorithm on this dataset we are able to 
corraborate the kink periodicity and associated decay time described by 
Verwichte~et~al.~\cite{Ver04}. However, in this instance no coronal-loop 
spatial periodicities could be found. The presented algorithm is therefore 
applicable to all datasets and is a reliable tool for detecting kink oscillations.

\begin{figure}
\epsscale{1.0}
\plotone{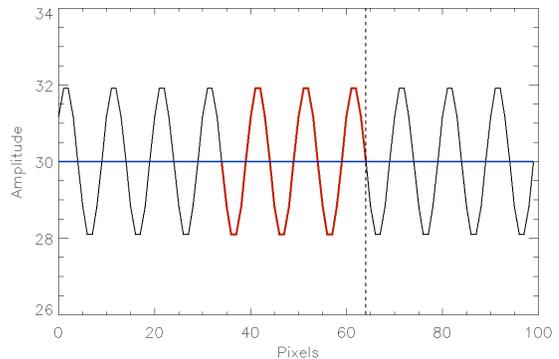}
\caption{The integrated wave form when the exposure time is 
equal to an integer number of complete periods. The black wave train is a sinusoid, equal 
in both amplitude and wavelength, to the observed {\sc{trace}} spatial oscillation. The bold red 
line indicates the portion, of the input wave train, which passes through an arbitrary 
spatial location, as denoted by the vertical black dashed line, during the exposure time. The result of integrating over the 
chosen exposure time is plotted as a bold blue line. It is clear that the resultant wave, when the 
exposure time is equal to an integer multiple of the period, equals zero. This wave will therefore 
not be detectable by the {\sc{trace}} instrument. \label{zero_sim}}
\end{figure}

\begin{figure}
\epsscale{1.0}
\plotone{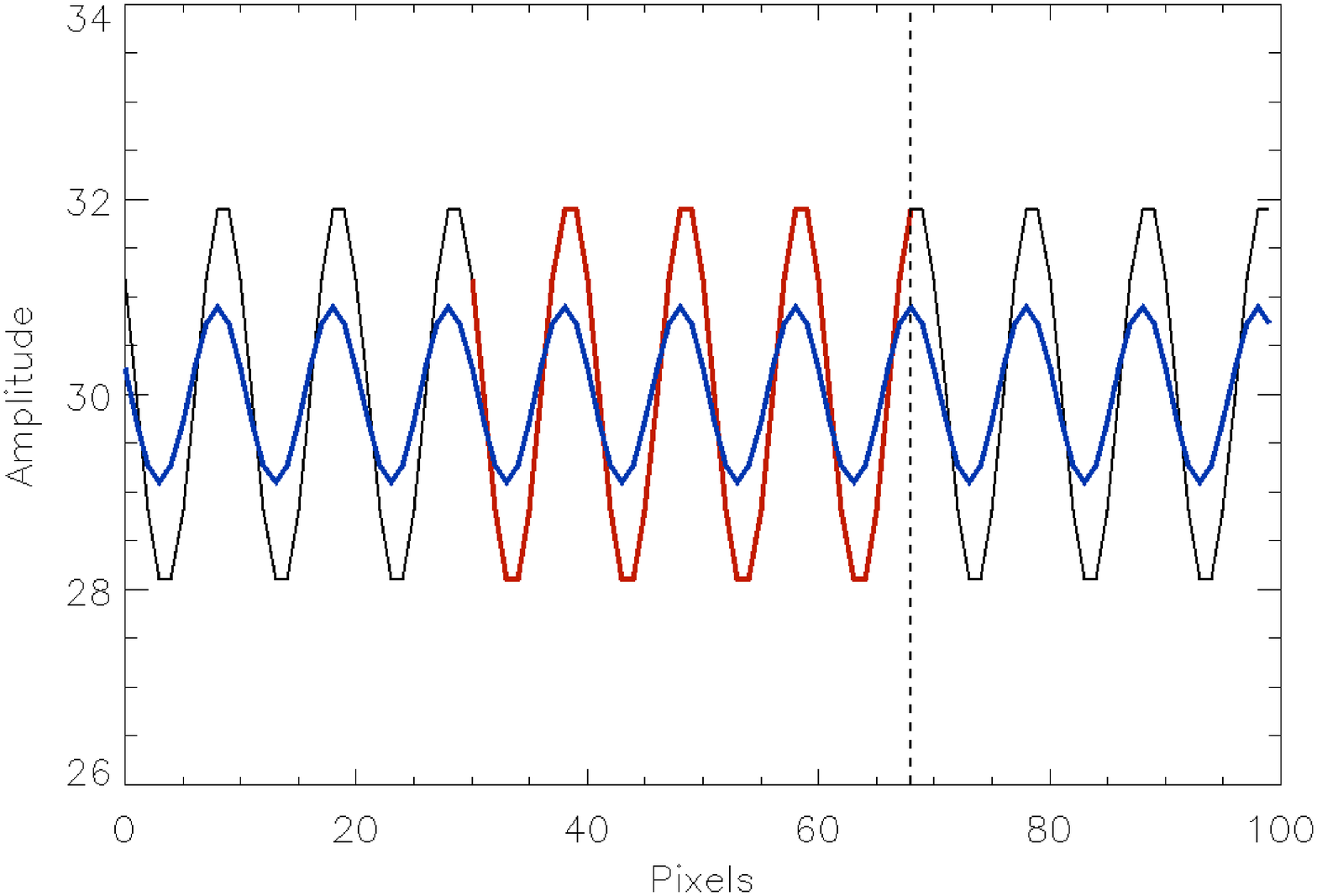}
\caption{Identical plot to Figure~\ref{zero_sim}, only showing the resultant wave form produced when 
the exposure time matches that of the {\sc{trace}} instrument. It is clear that the resultant wave, 
unmodified in wavelength, is still detectable even when multiple oscillations pass through the same 
pixel on the detector. This plot demonstrates that our detected 3500~km spatial periodicity is real. \label{true_sim}}
\end{figure}

Additionally, to verify that 
the detected oscillation is not an artifact of the loop-tracking algorithm, 
we have designed a numerical simulation to test the reliability of our oscillatory 
signal. A synthetic wave train is generated which is identical in both amplitude 
and wavelength to the detected {\sc{trace}} oscillations, and propagates 
with an Alfv\'{e}n speed of 1000~km/s (Nakariakov~\& Verwichte~2005). A resultant 
wave, formed from the numerical integration with time, is created based on real 
exposure times extracted from the {\sc{trace}} image files. As would be expected, 
if the exposure time is an exact integer multiple of the oscillatory period, the 
resultant wave form equals zero at all spatial locations (Fig~\ref{zero_sim}). This  
is due to,
\begin{equation}
\int_{0}^{n2\pi} \mathrm{Sin(}t\mathrm{)~d}t = 0
\end{equation}
where $n$ is an integer number and $t$ is time. By contrast, if the exposure time is 
set to a value {\it{not equal}} to an integer multiple of the period, then the 
propagating wave can be detected. Figure~\ref{true_sim} shows the resultant 
wave, formed using the exact exposure time taken from the {\sc{trace}} images. It is 
clear, from Figure~\ref{true_sim}, that even though multiple oscillatory cycles pass 
through the same pixel on the detector, an oscillatory periodicity can still be 
detected.

Throughout the analysis of the presented results, we interpret the translation of the 
coronal loop, both in the temporal {\it{and}} spatial domains, to be the 
result of the physical movement of this feature. Other aspects may accompany, and subsequently aid, 
in the movement of the loop. Things such as a change in temperature, or a change in inclination angle, 
may result in line-of-sight intensity changes. Since the {\sc{trace}} bandpass used allows 
a wide range of coronal temperatures to be investigated, and since the movement of the loop edge is 
restricted to $<10''$, we feel confident that the main source of translation is caused by the physical 
movement of the loop itself.

~

~

\section{Concluding Remarks}
\label{conc}

A new method for the detection, and subsequent tracking, of 
coronal loop structures is presented. This method also provides the 
ability to simultaneously search for longitudinal and transverse 
oscillations in real-time. It is possible to detect the periodic 
loop displacement caused by the influence of transverse waves, and in 
addition the intensity fluctuations, due to density restrictions and 
rarefactions, synonymous with longitudinal 
oscillations can also be easily identified through concurrent time-series 
analysis. Through the subtraction of loop outlines,  
we are essentially removing general features and shapes which remain  
constant from frame to frame, yet magnifying features which evolve or change 
spatial position.  
%This is particularly useful for future
%space-borne solar telescopes, where the large achievable data rates
%will make real-time data analysis paramount. 
Utilizing our method 
for automated coronal loop analysis, we have detected spatial 
periodicities, on the scale of 3500~km per cycle, occurring on a 
coronal loop immediately after being buffeted by a neighbouring M4.6 
flare. We detect the damping of these loop-top oscillations, with a 
45\% reduction in oscillatory power over a 222~s interval. In addition, we find 
a wrapped spatial phase shift of $0^{\circ}$ to 
$-50^{\circ}$, related to the time interval of 222~s between 
successive kink-oscillation equilibrium positions, for the 
travelling loop-top oscillations.

We propose the above methodology will be useful in the
future, especially with both temporal and spatial resolutions of
space-borne solar telescopes constantly improving. From coronal
imagers available on the upcoming {\it{Solar Dynamics Observatory}}
and {\it{Solar Orbiter}}, as well as the recently launched
{\it{Hinode}} satellite, there will be a wealth of high-resolution
images containing abundances of coronal loop oscillations. There are
a number of currently available methods to detect and analysis such
oscillatory phenomena. Such techniques may be data specific, and therefore 
not applicable to a wide range of varying coronal observations. However, our 
technique is designed to be readily applicable to all coronal datasets, and 
is a great asset when spatial resolutions are of the highest order due to the 
WTMM techniques outlined here. We suggest that the currently
presented method provide further excellent capabilities for the
detection of both temporal {\it and} spatial coronal loop
oscillations.

\acknowledgments

DBJ is supported by a Northern Ireland Department for Employment and 
Learning studentship. DBJ additionally thanks NASA Goddard Space 
Flight Center for a CAST studentship -- in particular Doug Rabin and 
Roger Thomas deserve special thanks for their endless help, support 
and scientific input. RE acknowledges M. K\'eray for 
patient encouragement and is also grateful to NSF, Hungary (OTKA, 
Ref. No. K67746). FPK is grateful to AWE Aldermaston for the 
award of a William Penney Fellowship. Wavelet software was 
provided by C. Torrence and G.P. Compo.\footnote{Wavelet software 
is available at http://paos.colorado.edu/research/wavelets/.}

\clearpage

%\begin{figure}
%\epsscale{1.0}
%\plotone{Figures/filtered.eps}
%\caption{Original {\sc{trace}} image (left) and resulting fine-structure image after undergoing Lapacian convolution (right). A false
%colour table is used to show the fine detail detected through use of the Laplacian convolution. In both images the axis scales are
%pixel number with $1$~pixel~$=~0.5\arcsec$. \label{filtered}}
%\end{figure}

%\begin{figure}
%\epsscale{1.0}
%\plotone{Figures/sharpened.eps}
%\caption{Original {\sc{trace}} image (left) and resulting image after undergoing full image sharpening (right). It is clear to see that
%fine structures have been emphasized during the Laplacian filtration. In both images the axis scales are
%pixel number with $1$~pixel~$=~0.5\arcsec$. \label{sharpened}}
%\end{figure}

%\clearpage

\clearpage

\clearpage

\clearpage

\clearpage

%\begin{figure}
%\epsscale{0.9}
%\plotone{/scratch/dbj/Thesis/papers/edge_paper/Figures/subtracted_control.eps}
%\caption{Identical diagram to Figure~\ref{subtracted_flare} except for the use of a non-oscillating control loop. Notice how no
%spatial periodicities are detected in this example. \label{subtracted_control}}
%\end{figure}

\clearpage

\clearpage

\end{document}